# Modeling Irregular Boundaries Using Isoparametric Elements in Material Point Method

**Ezra Y. S. Tjung, P.E., S.M.ASCE[1], Shyamini Kularathna, Ph.D[2], Krishna Kumar, Ph.D[3] and Kenichi Soga, Ph.D., M.ASCE[4]**

[1]Department of Civil and Environmental Engineering, University of California, Berkeley, Berkeley, CA 94720-1234; e-mail: ezrayst@berkeley.edu

[2]Department of Civil and Environmental Engineering, University of California, Berkeley, Berkeley, CA 94720-1234; e-mail: kshyamini@berkeley.edu

[3]Department of Civil, Architectural and Environmental Engineering, The University of Texas at Austin, Austin, TX 78712; e-mail: krishnak@utexas.edu

[4]Department of Civil and Environmental Engineering, University of California, Berkeley, Berkeley, CA 94720-1234; e-mail: soga@berkeley.edu

**ABSTRACT**

The Material Point Method (MPM) is a hybrid Eulerian-Lagrangian approach capable of simulating large deformation problems of history-dependent materials. While the MPM can represent complex and evolving material domains by using Lagrangian points, boundary conditions are often applied to the Eulerian nodes of the background mesh nodes. Hence, the use of a structured mesh may become prohibitively restrictive for modeling complex boundaries such as a landslide topography. We study the suitability of unstructured background mesh with isoparametric elements to model irregular boundaries in the MPM. An inverse mapping algorithm is used to transform the material points from the global coordinates to the local natural coordinates. Dirichlet velocity and frictional boundary conditions are applied in the local coordinate system at each boundary node. This approach of modeling complex boundary conditions is validated by modeling the dynamics of a gravity-driven rigid block sliding on an inclined plane. This method is later applied to a flume test of controlled debris flow on an inclined plane conducted by the United States Geological Survey (USGS).

**INTRODUCTION**

The Material Point Method (MPM) is a continuum approach for modeling large-deformations in history-dependent materials (Sulsky et al., 1994 and 1995). Unlike the classical mesh-based numerical approaches such as the Finite Element Method (FEM), the MPM avoids mesh distortion issues when solving large-deformation problems. The material domain is discretized into a set of Lagrangian material points, which carry information such as their mass, velocity and other history-dependent material variables. The Eulerian background mesh is used purely for solving partial differential equations and the material points can traverse independent of this background mesh. An illustration of the MPM algorithm is shown in Figure 1.

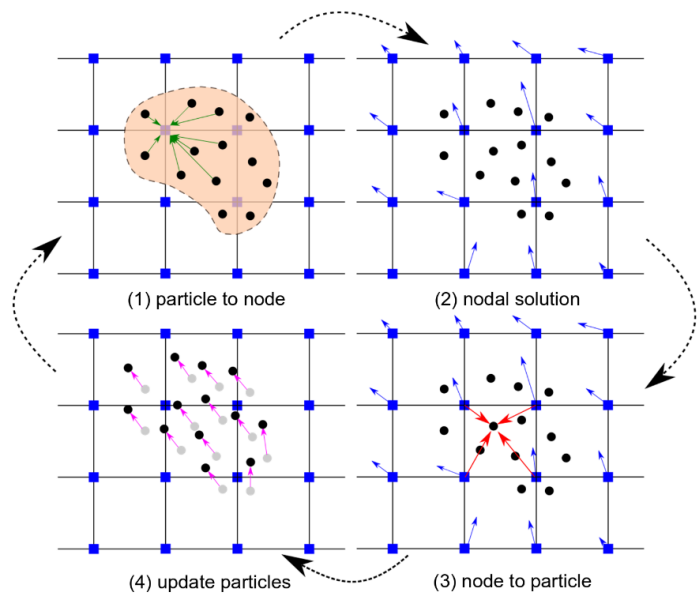

**Figure 1: Illustration of the MPM algorithm (1) A representation of material points overlaid on a background computational mesh. Green arrows represent material point state vectors (mass, volume, velocity, etc.) being projected to the nodes of the computational mesh. (2) The equations of motion are solved onto the nodes, resulting in updated nodal velocities and positions. (3) The updated nodal kinematics is interpolated back to the material points. (4) The location of the material points is updated, and the computational mesh is reset (reproduced after Soga et al., 2016)**

Typically in the MPM, rectilinear elements are used to represent a structured mesh. Kinematic boundary constraints are applied on the Eulerian background nodes, independent of the location of the Lagrangian material points. Hence, modeling irregular boundaries such as a natural landslide topography remains a challenge. Figure 2 presents the commonly adopted methods in the MPM to model irregular boundaries in the rectilinear structured mesh. Figure 2a illustrates the application of Dirichlet velocity constraints on the node closest to the boundary. Xu et al. (2018) adopted this to model the 3D runout of Hongshiyan landslide in China. This approach of constraining the boundary nodes often results in a step-wise boundary constraint that does not accurately capture the complex boundaries observed and may cause unrealistic flow constraints near the boundary. Another commonly adopted approach is to represent the irregular boundary surface using a contact algorithm between distinct sets of material points as shown in Figure 2b. Various landslides simulations have utilized this approach such as the Oso landslide simulations by Yerro et al. (2018). This approach is typically used with an unstructured triangular or tetrahedral elements. Figure 2c illustrates a nonconforming implicit boundary condition that can be manipulated independently of the background mesh. In this approach, the Dirichlet boundary conditions are weakly imposed on the finite element space. Cortis et al. (2018) first applied an implicit boundary method in MPM by weakly imposing the Dirichlet boundary conditions over a finite thickness along the boundary. Nevertheless, more research is needed to extend this method for practical applications of MPM.

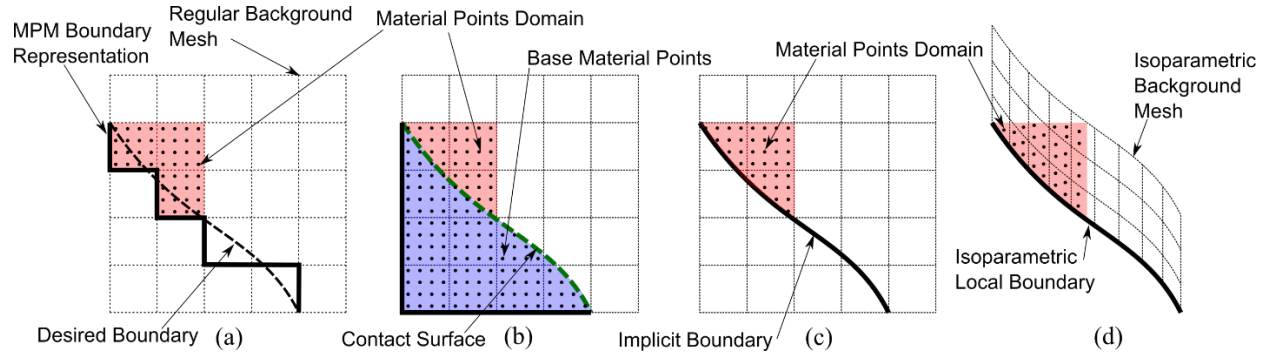

Figure 2: Illustration showing four different methods to model irregular boundaries in MPM: (a) use the closest element to represent boundary conditions, (b) use material points as the topography with the contact surface, (c) use implicit boundary, and (d) use isoparametric elements to model the topography.

In this study, we explore the suitability of unstructured quadrilateral elements that conform to irregular boundaries as illustrated in Figure 2d. The use of isoparametric representation in the MPM has the following implications: (1) the constraints on boundary surfaces represented using irregular isoparametric elements are often not aligned with the axes of the global coordinate system and have to be defined in the local coordinate system; (2) in the MPM, nodal properties such as the mass, momentum and forces are mapped from the material points based on their location in the natural coordinate system; although mapping from the natural coordinate $(\xi, \eta)$ to the Cartesian coordinate $(x, y)$ is a linear transformation, the inverse mapping is not generally defined for irregular elements. The approach adopted in this study to solve the above two issues are discussed in the following sections. The method is then validated against the analytical solution of a gravity-driven rigid block sliding on an inclined plane and is applied to the simulation of dry sand flow in a controlled experimental inclined flume (Denlinger and Iverson 2001).

**IRREGULAR BOUNDARY CONDITIONS**
In the MPM with a rectilinear Cartesian background mesh, the Dirichlet velocity and frictional constraints are applied on the background nodes, in the directions aligned with the global coordinate axes. However, irregular topologies such as those of a landslide have boundary constraints (velocity vector components) not aligned with the global coordinate axes, and therefore should be applied in the local coordinate axes of each node. This requires a transformation of the nodal velocity and acceleration vectors from the global coordinate system to the local coordinate system through a transformation matrix: $v_L = T v_G$, where $T$ is the transformation matrix, $v_G$ and $v_L$ are the velocity vectors in the global and the local coordinates, respectively.

The following steps are involved when applying nodal constraints for irregular elements: (1) compute the transformation matrix for each constrained node, (2) transform nodal velocity and acceleration vectors to the local coordinate system using the transformation matrix, (3) apply the Dirichlet velocity and frictional boundary constraints in the local coordinate system, and (4) transform the nodal velocity and acceleration vectors from the local coordinate system back to the global coordinate system.

**INVERSE MAPPING IN ISOPARAMETRIC ELEMENTS**

In contrast to the Gauss integration used in the FEM, the MPM considers the locations of material points as the integration points. This converts the integrals over the momentum equation into summation over the material points in a semi-discrete formulation. The properties at the material points such as its mass, momentum, and stresses are mapped to the corresponding nodal mass, momentum, and forces using the shape functions defined in the natural coordinate system. This process requires computing the natural coordinates of the material points in a reference element at each time-step. Figure 3 illustrates the inverse mapping of a point $p(x,y)$ in the global Cartesian coordinate system to the same point $p(\xi, \eta)$ in the natural coordinate system in a reference element. Although this inverse mapping is a straightforward linear scaling for rectilinear elements aligned with the global Cartesian coordinate axes, it is not well-defined for irregular elements.

This study adopts a set of general solutions for the inverse mapping of linear quadrilateral elements presented by Zhao et al. (1999) and can be written in a general matrix form:

$$[b_1\ c_1\ b_2\ c_2][\xi\ \eta] = [d_1 - a_1\xi\eta\ \ d_2 - a_2\xi\eta], \tag{1}$$

where,

$$\begin{aligned} d_1 &= 4x - (x_1 + x_2 + x_3 + x_4), \\ d_2 &= 4y - (y_1 + y_2 + y_3 + y_4), \end{aligned} \tag{2}$$

and,

$$[a_1\ a_2\ b_1\ b_2\ c_1\ c_2] = [1\ -1\ 1\ -1\ -1\ 1\ 1\ -1\ -1\ -1\ 1\ 1\ 1][x_1\ y_1\ x_2 \tag{3}$$

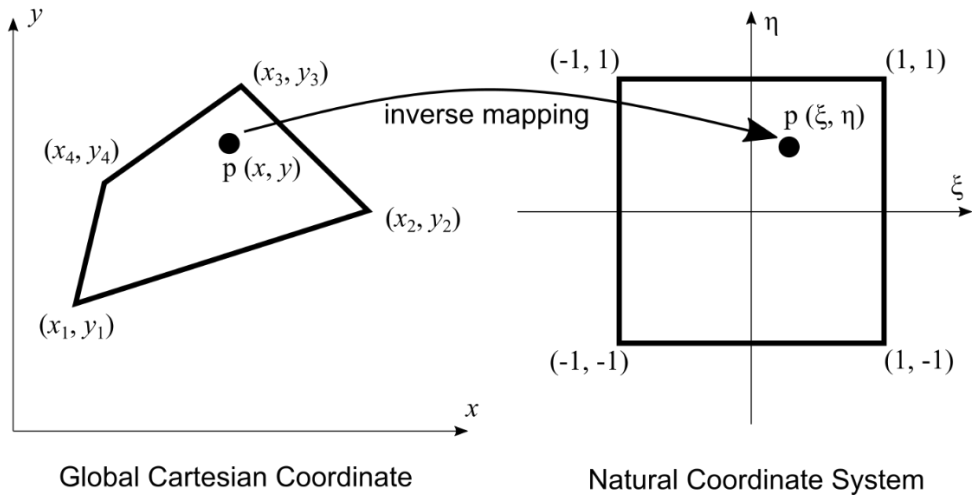

**Figure 3: Inverse mapping of point *p* from a quadrilateral isoparametric element in the global Cartesian coordinate system to the natural coordinate system.**

The approach of modeling irregular boundaries using the isoparametric elements is implemented in the CB-Geo MPM code (Kumar et al., 2019).

**NUMERICAL EXAMPLES**
**Rigid Block**
This approach of modeling irregular boundaries is validated by simulating a gravity-driven rigid block sliding down a 30-degree inclined plane. The background mesh consists of quadratic elements of size 0.25 by 0.25 m. The rigid block is modeled as a linear elastic material with a density of 1800 kg/m$^3$, a very high Young's modulus of 1 GPa and zero Poisson's ratio. An acceleration of -9.81 m/s$^2$ due to gravity is applied in the vertical y-direction. The rigid block is discretized with 256 material points with 4-by-4 material points located in each element at the initial state. Figure 4 illustrates the evolution of the locations and velocities of the rigid block as it descends down the inclined slope under gravity with time. All material points in the block have a uniform velocity as seen in Figure 4. Figure 5 shows that the MPM simulations are able to replicate the evolution of velocities of a rigid block sliding down an inclined plane against and the analytical solution for different basal friction coefficients. The MPM is able to accurately capture the velocities of the material points with an average error of less than 0.01% at each time step.

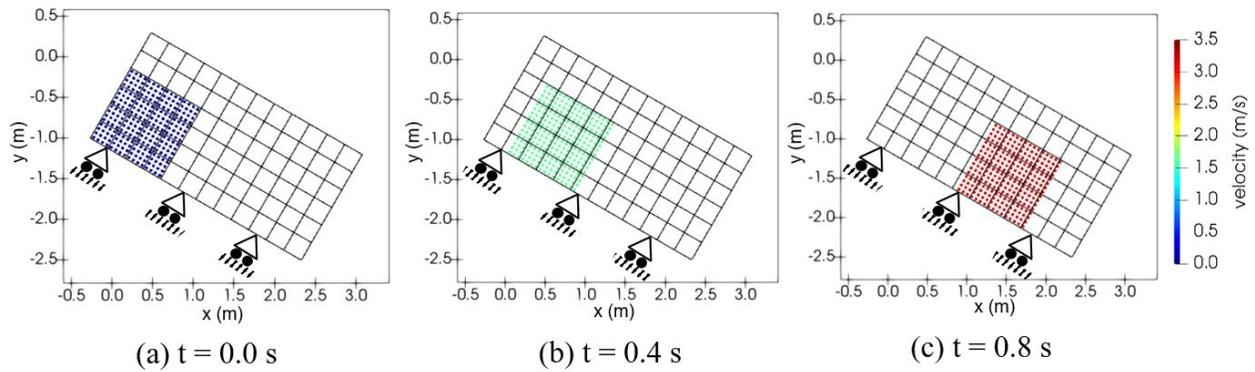

(a) t = 0.0 s    (b) t = 0.4 s    (c) t = 0.8 s

**Figure 4: Illustration of the evolution of the locations and the velocity magnitude of the rigid block sliding down an inclined plane with a friction coefficient of 0.1.**

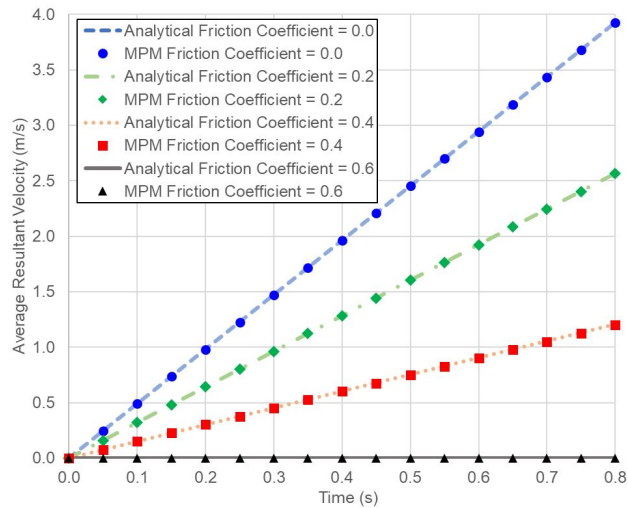

**Figure 5: The time evolution of the average velocity magnitude of the material points for different basal friction coefficients**

**Flume Test**

The capabilities of this MPM framework for modeling irregular boundaries are further demonstrated by modeling the flow of dry sand in a rectangular flume conducted by the United States Geological Survey (USGS). Figure 6a illustrates the geometry of the experimental flume used by Denlinger and Iverson (2001). The flume is approximately 2 m long with a bed surface inclined at 31.4° adjoined to a horizontal runout surface by a curved section with a radius of curvature of 0.1 m. Loose-packed, well-sorted and well-rounded dry quartz sand with an average grain diameter of about 0.5 mm is positioned approximately 37.5 cm upslope from the break in slope. The sand is released en masse, almost instantaneously, by a spring-loaded gate.

A 2D MPM plane strain simulation of this experiment is performed using an unstructured mesh of irregular linear quadrilateral elements. The sand is discretized into 5,254 material points with constant volume. At the start of the simulation, the material points are assigned an initial stress corresponding to the geo-static condition. Figure 6b presents the initial configuration of the simulation, showing the background mesh along with the initial locations of the material points.

The mesh consists of 17,600 isoparametric elements. The velocity and frictional boundary conditions are applied on the bottom nodes in the direction normal to the sliding plane as shown in Figure 6b. Further details about the mesh and the material points parameters are summarized in Table 1.

**Table 1: Simulation parameters for the flume test**

|          | Parameters                          | Symbol            | Values              |
|----------|-------------------------------------|-------------------|---------------------|
| Model    | Number of material points           | $N_{mp}$          | 5,254               |
|          | Average material points spacing     | $l_{mp}$ (mm)     | 0.625               |
|          | Number of elements                  | $N_e$             | 17,600              |
|          | Average element spacing             | $L_e$ (mm)        | 2.5                 |
|          | Gravitational acceleration          | $g$ (m/s²)        | -9.81 (y-dir)       |
| General  | MPM algorithm                       | -                 | Update Stress First |
|          | Total # of steps                    | $t_{steps}$       | $1.5 \times 10^6$   |
|          | Time step                           | $\Delta t$ (s)    | $1.0 \times 10^{-6}$ |
|          | The total duration of the simulation| $t_{final}$ (s)   | 2.0                 |
| Material | Material model                      | -                 | Mohr-Coulomb        |
|          | Young's modulus                     | $E$ (Pa)          | $2.0 \times 10^6$   |
|          | Poisson's ratio                     | $\nu$             | 0.3                 |
|          | Internal friction angle             | $\phi'$ (°)       | 40                  |
|          | Dilation angle                      | $\psi'$ (°)       | 0                   |
|          | Cohesion                            | $c'$ (Pa)         | 0                   |
|          | Tension cutoff                      | $\sigma_t$ (Pa)   | 0                   |
|          | Density                             | $\rho$ (kg/m³)    | 1,600               |
| Boundary | Basal friction coefficient          | $\mu$             | 0.52                |

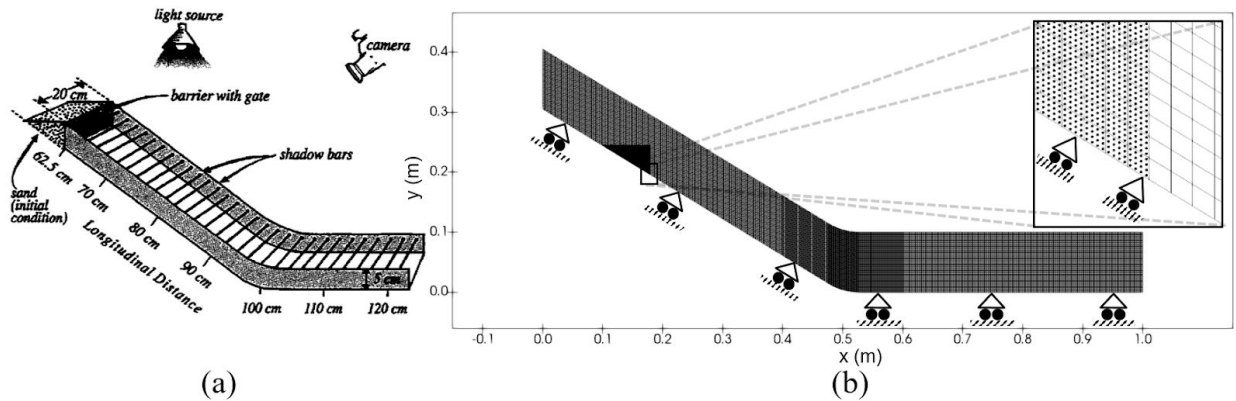

(a)                                             (b)

**Figure 6: (a) Schematic figure of the dry sand flow in miniature flume experiment reproduced after Delinger and Iverson (2001), (b) initial state of the 2D MPM model.**

The results of the simulation are compared with experimental results along the centerline of the flume. Figure 7 compares the evolution of run-out with time between the experiment and the MPM simulation. The MPM is able to capture the flow pattern observed in the experiments in the initial 0.5 s. In this initial stage, both the locations of the front and tail of the flow in the simulation are in good agreement with the experiment. Beyond 0.5 s, the flow front observed in the MPM is very close to the experiment results, while the tail continues to fall behind the experimental observations.

Figure 8 compares the evolution of the front and tail of the flow between the experiment and the MPM simulation. The evolution of the flow front in the MPM simulation is in line with the experiment results, with the maximum difference of only 0.5%. However, the tail of the flow only matches in the first 0.5 s, beyond which the MPM simulations predict a longer tail in contrast to the experiments. This behavior can be attributed to the lack of momentum in the material points located at the trailing part of the flow. The fewer the number of material points representing the trailing part of granular flow contributes to the mismatch between the MPM simulation and the experiment. The velocities of the material points in the bottom elements are affected by the nodal constraints. Reducing the mesh size and increasing the number of material points representing the granular media would improve the results of the simulation.

Figure 9 shows the evolution of the average velocity magnitude of all the material points with time. As the soil mass slides down the slope, it gains momentum due to gravity and thus increasing the velocity of the material points representing them. They reach an average peak velocity magnitude of 0.45 m/s at 0.6 s. Then, as the soil mass slide past the curved section, the velocity gradually decreases until it levels off at 1.55 s. Beyond this, there is only minor movement. This is consistent with the observation by Delinger and Iverson (2001) that the deposition is complete 1.5 s after the flow release.

Figure 10 shows the final configuration of the material points with respect to their initial configurations. The material points starting at the front end up in front of and underlying the other material points starting behind them. The material points show a domino-type toppling behavior in the run-out evolution. The thickest deposition is observed near the curved section of the flume. Majority of the granular media found on the horizontal plane is from the right-most one-quarter of the volume (blue colored material points). This implies that the subsequent arrival of granular mass did not push the flow front further along the flume, an observation noted by Delinger and Iverson (2001).

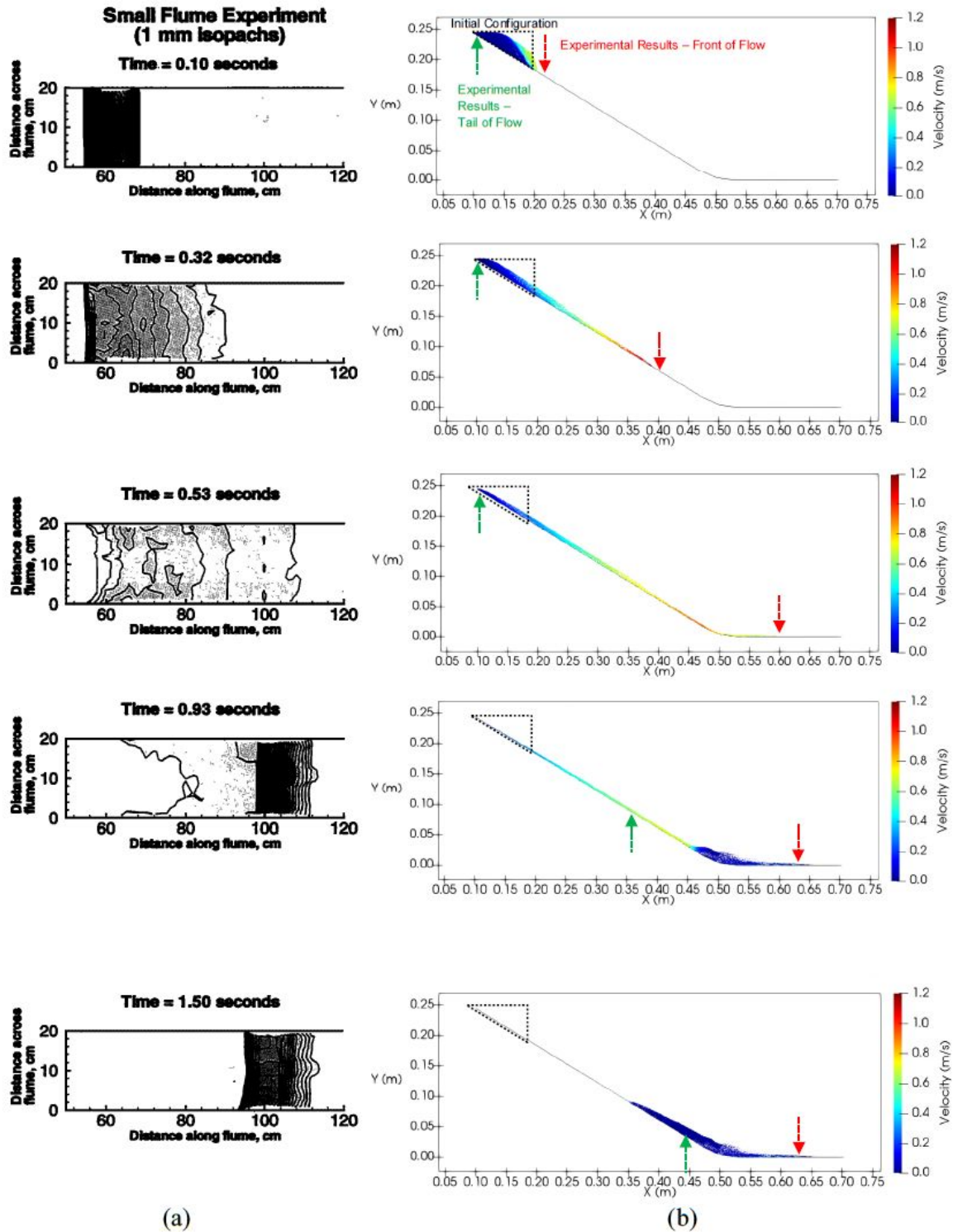

(a)                                                                                       (b)

Figure 7: Evolution of dry sand flow with time: (a) experimental results (reproduced after Delinger and Iverson, 2001), (b) the MPM simulation

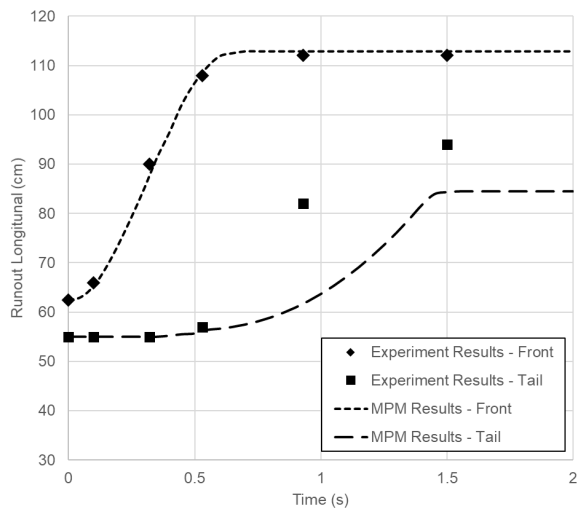

Figure 8: Run-out evolution of dry sand flow with time: (a) experimental results from Delinger and Iverson (2001), and (b) MPM results

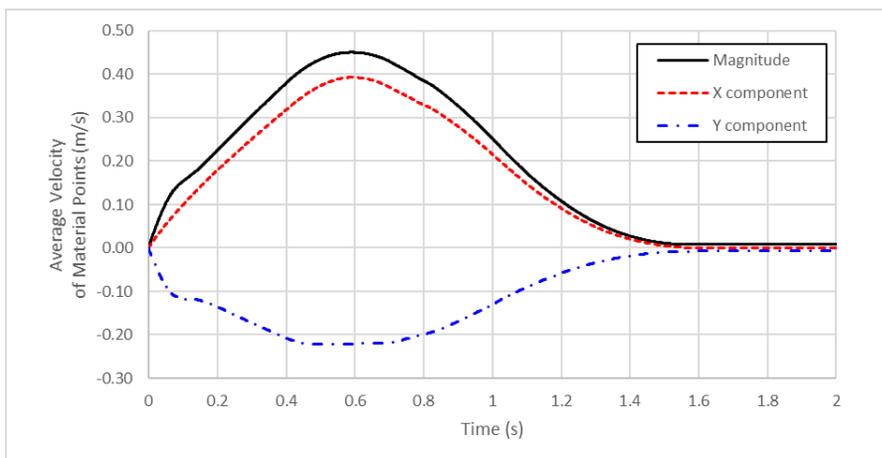

Figure 9: The evolution of the velocity components and magnitude of all the material points in MPM simulation with time.

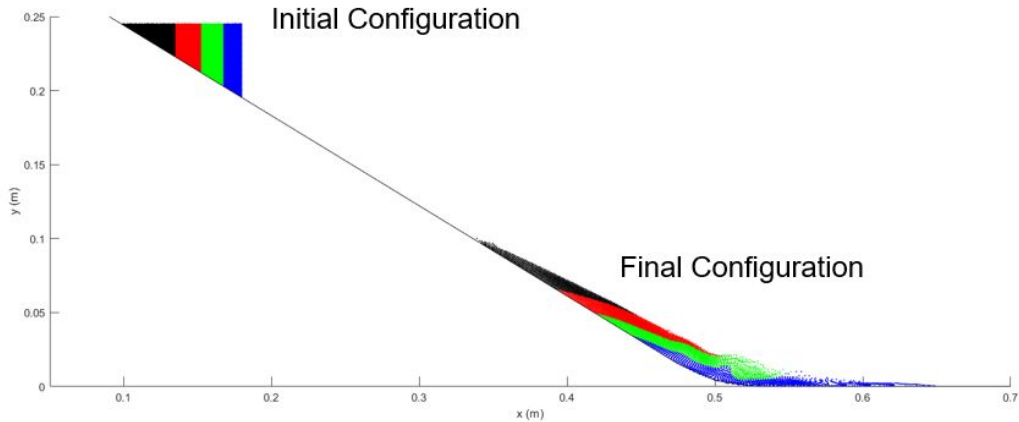

**Figure 10: Location of material points at the initial and the final configuration**

This approach of treating irregular boundary conditions in the MPM has been successfully applied to model the experimental flow of dry granular sand and capture the general flow dynamics observed by Denlinger and Iverson (2001).

**SUMMARY**

In this paper, we proposed the use of conforming mesh with irregular quadrilateral elements to model complex boundary surfaces in MPM. The nodal velocity and acceleration variables are required to be transformed into its local coordinate system to apply the kinematic constraints. Moreover, a non-linear inverse mapping algorithm is implemented to map the material points onto the reference quadrilateral element coordinate system. We then validate the approach against the analytical solution of a simple gravity-driven block moving along an inclined plane and a granular flow in a flume experiment. This technique could be adopted to model irregular topography such as a landslide.